\documentclass[preprint,12pt]{aastex}
\usepackage{latexsym}

\shorttitle{DEM\,L\,106}
\shortauthors{Chu, Chen, Danforth, Dunne, Gruendl, Naze, Oey, Points}

\newcommand {\dem}{DEM\,L\,106}
\newcommand {\ha}{H$\alpha$}

\newcommand {\nii}{[\ion{N}{2}]}
\newcommand {\hii}{\ion{H}{2}}
\newcommand {\kms}{km~s$^{-1}$}

\begin{document}

\title{The Wind of the B[e] Supergiant Hen S22 Viewed through a
 Reflection Nebula in DEM\,L\,106}

\author{You-Hua Chu\altaffilmark{1,}\altaffilmark{2}, C.-H. Rosie Chen\altaffilmark{1}, 
Charles Danforth\altaffilmark{2,}\altaffilmark{3}, Bryan C.\ Dunne\altaffilmark{1}, 
Robert A.\ Gruendl\altaffilmark{1}, Ya\"el Naz\'e\altaffilmark{4,}\altaffilmark{5}, 
M.~S.\ Oey\altaffilmark{6}, Sean D.\ Points\altaffilmark{2,}\altaffilmark{7}}
\altaffiltext{1}{Astronomy Department, University of Illinois, 
        1002 W. Green Street, Urbana, IL 61801;
        chu@astro.uiuc.edu, c-chen@astro.uiuc.edu, carolan@astro.uiuc.edu, 
        gruendl@astro.uiuc.edu}
\altaffiltext{2}{Visiting astronomer, Cerro Tololo Inter-American Observatory}
\altaffiltext{3}{Department of Physics and Astronomy, Johns Hopkins University, 
        3400 N. Charles Street, Baltimore, MD 21218; danforth@pha.jhu.edu}
\altaffiltext{4}{Institut d'Astrophysique et de G\'eophysique, All\'ee du 6
        Ao\^ut 17, Bat. B5c, B 4000 Li\`ege (Sart-Tilman), Belgium; 
        naze@astro.ulg.ac.be}
\altaffiltext{5}{Research Fellow FNRS (Belgium)}
\altaffiltext{6}{Lowell Observatory, 1400 W. Mars Hill Rd., Flagstaff, AZ 86001; 
        msoey@lowell.edu}
\altaffiltext{7}{Department of Physics and Astronomy, Northwestern University, 
       2145 Sheridan Rd., Evanston, IL 60208; s-points@northwestern.edu}

\begin{abstract}

Narrow-band $HST$ WFPC2 images reveal a bow-shock-like halo around
the \hii\ region N30B toward the B[e] supergiant Hen S22 located 
within the larger DEM\,L\,106 nebula in the Large Magellanic Cloud.  
High-dispersion spectra of N30B show a narrow \ha\ 
emission component from the ionized gas; the velocity 
variations indicate a gas flow of $-$5 to $-$10 \kms\
in the vicinity of the \hii\ regions, which is resultant
from interactions with Hen S22's stellar wind and responsible 
for the bow-shock morphology.  
Spectra of N30B's halo show broad \ha\ profiles extending
over $>1000$ \kms, similar to that of Hen S22, indicating that 
the halo is a reflection nebula of Hen S22.  
Broad-band morphologies of N30B's halo are also consistent 
with the reflection nebula interpretation.
We use dust-scattering properties and the observed brightnesses 
of the reflection nebula and Hen S22 to constrain the reflection 
geometry.  
The reflected stellar \ha\ emission and absorption vary across 
the reflection nebula as a result of viewing S22's anisotropic 
wind across varying angles.  This reflection nebula, together 
with the edge-on orientation of Hen S22's disk, provides an 
invaluable opportunity to study the disk and polar winds of 
a B[e] supergiant.

\end{abstract}


\section{Introduction}

A {\it Hubble Space Telescope (HST)} WFPC2 \ha\ image of the \hii\ 
region \dem\ \citep{DEM} in the Large Magellanic Cloud (LMC) revealed 
a pair of bright, compact \hii\ regions embedded within a large 
shell nebula \citep{chenetal00}.  \dem\ has been previously cataloged 
by \citet{H56} as N30, and the two compact \hii\ regions were 
identified as an \hii\ knot and designated as N30B.  As shown
in Figure~1, while the main 
body of N30B is unremarkable, it is surrounded by a halo with sharp 
edges on the northern rim facing the luminous B[e] supergiant Hen S22 
\citep[= HD\,34664;][]{H56}.  B[e] supergiants are known to possess
stellar winds \citep{Zetal96}.  The nebular morphology and 
relative location of Hen S22 and N30B thus appear to suggest 
that the halo of N30B is a bow shock produced by dynamical 
interactions with the wind of Hen S22. 

A two-component stellar wind model has been proposed for B[e] supergiants:
(1) a cool, slow wind along a dusty equatorial disk which gives 
rise to narrow, low-excitation emission lines of, e.g., \ion{Fe}{2},
[\ion{Fe}{2}], and [\ion{O}{1}], and (2) a hot, fast polar wind that 
is responsible for the broad, high-excitation absorption lines of,
e.g., \ion{C}{4}, \ion{Si}{4}, and \ion{N}{5} \citep{Zetal85,Zetal86}.  
$IUE$ spectra of Hen S22 in the 1800--3200 \AA\ range are dominated
by P Cygni type lines of \ion{Fe}{2}, from which it is derived that
the equatorial disk is viewed nearly edge-on and the terminal wind 
velocity along the disk is $\sim80$ \kms \citep{Zetal96}; while $IUE$ 
spectra in the 1150--1950 \AA\ range show blue-shifted absorption of 
\ion{Si}{4} and \ion{C}{4} indicating that the polar wind of Hen S22 
has an expansion velocity $>$1000 \kms\ \citep{Betal83,Zetal86}.
N30B is projected at 20$''$--30$''$ \citep[or 5--7.5 pc for a distance 
of 50 kpc,][]{F99} from Hen S22; its 24$''$ spatial extent (Figure 1) 
has a good chance to be partially 
located above the disk and interact with the polar wind of Hen S22.

To determine the nature and physical properties of the bow-shock-like
halo around N30B, we have obtained high-dispersion, long-slit echelle 
spectra of N30B, its bow-shock-like halo, its surrounding interstellar 
medium, and Hen S22.  We find that the bow-shock-like halo shows 
surprisingly broad \ha\ line profiles similar to that of Hen S22, 
suggesting that it is a reflection nebula.  
Furthermore, the spectral features vary from the east side
to the west side of N30B, indicating that {\it N30B indeed sees the wind
of Hen S22 along a range of viewing angles}. 
Recently archived $HST$ WFPC2 images of N30B obtained by the 
Hubble Heritage Project provide additional support to the 
interpretation of a reflection nebula for the halo of N30B.
In this paper, we report these observations, and discuss the physical 
properties of N30B, the nature of its halo, and the implication of 
our results on the disk and polar winds of the B[e] supergiant Hen S22.

\section{Observations and Data Reduction}

\subsection{HST WFPC2 Images}

Our $HST$ WFPC2 images of DEM\,L\,106 (PI: Chu) were obtained on 1998 
November 14.  The observations were made through the {\it F656N} (\ha)
filter for $3\times800$ s.  The images were processed using the 
standard $HST$ data pipeline and then combined to remove cosmic rays 
and to produce an image with a total exposure time of 2400 s.  
We further processed the resultant image with the standard procedures
outlined in \citet{chenetal00} to extract \ha\ fluxes.

Two corrections have been applied to these \ha\ fluxes
due to the $\sim$300 km~s$^{-1}$ red-shift of the LMC.
First, the filter transmission of the red-shifted \ha\ line is 
$\sim93$\% of the peak transmission, so the extracted \ha\
fluxes are multiplied by a correction factor of 1.07.
Second, the \nii\ $\lambda$6548 line is red-shifted 
into the \ha\ passband at $\sim90$\% of the peak transmission.
Using the \nii\ $\lambda$6548/\ha\ ratio measured from our
echelle spectra (see \S2.2), we find that the \nii\ $\lambda$6548
line contributes to $\sim4$\% of the \ha\ flux.  Therefore,
an additional correction factor of 0.96 is applied to the 
extracted \ha\ fluxes.

The $HST$ WFPC2 images of DEM\,L\,106 for the Hubble Heritage 
Project (PI: Noll) were obtained on 2001 October 18.
The observations were made with the $F439W$ (WFPC2 $B$),
$F555W$ (WFPC2 $V$), $F814W$ (WFPC2 $I$), $F502N$ ([\ion{O}{3}]), 
and $F673N$ ([\ion{S}{2}]) filters for exposure times of 
$2\times160$, $2\times40$, $2\times100$, $2\times500$, and 
$2\times1100$ s, respectively.
Note that the broad-band images may contain contributions from
nebular lines, e.g., the $B$ band contains the H$\gamma$, 
H$\delta$, and H$\epsilon$ lines and the $V$ band contains the
H$\beta$ line.
We have processed these images using standard procedures.
We have also adjusted the coordinates using astrometric solutions
calculated from the GSC2.2\footnote{The Guide Star Catalogue-II is 
a joint project of the Space Telescope
Science Institute and the Osservatorio Astronomico di Torino. Space
Telescope Science Institute is operated by the Association of 
Universities for Research in Astronomy for the National Aeronautics 
and Space Administration under contract NAS5-26555. The participation of the
Osservatorio Astronomico di Torino is supported by the Italian Council for
Research in Astronomy. Additional support is provided by European Southern
Observatory, Space Telescope European Coordinating Facility, the
International GEMINI project and the European Space Agency Astrophysics
Division.} 
positions of 9 stars on the WF2 chip;
our final coordinates should have $\sim$ 0\farcs2 accuracy.
The WFPC2 images of N30B are presented in Figure 2.

\subsection{High-Dispersion Echelle Spectroscopy}

High-dispersion spectra of DEM\,L\,106 were obtained with the echelle
spectrograph on the 4 m telescope at Cerro Tololo Inter-American
Observatory (CTIO) from two observing runs, 2000 January 22 and
December 7.  Both runs used the 79 line mm$^{-1}$ echelle
grating in the single-order, long-slit observing configuration,
in which a flat mirror replaced a cross disperser and a post-slit
\ha\ filter was inserted to isolate a single order.  The red long 
focus camera and the 2000$\times$2000 SITe2K\_6 CCD were used to 
record data.  The 24 $\mu$m pixel size corresponds to roughly 
0.082 \AA\ along the dispersion and 0\farcs26 perpendicular 
to the dispersion.  Limited by the optics of the camera, the useful 
spatial coverage is $\sim$3$'$.  The spectral coverage is wide 
enough to include both the \ha\ line and the flanking 
\nii~$\lambda$$\lambda$6548, 6583 lines.  A slitwidth of 250 $\mu$m 
(1\farcs64) was used and the resultant FWHM of the instrumental profile 
was $13.5\pm0.5$ km s$^{-1}$.  The spectral dispersion was calibrated 
by a Th-Ar lamp exposure taken in the beginning of the night,  but the 
absolute wavelength was calibrated against the geocoronal \ha\ line 
present in the nebular observations.  The journal of echelle 
observations is given in Table~1.

\section{Discussion}

The two small \hii\ regions in N30B are aligned roughly along the 
east-west direction, thus we call these two regions N30B-E and 
N30B-W, respectively.  
In the \ha\ image, apart from the bow-shock-like halo, the bright
central regions of these two \hii\ regions have simple morphologies 
expected for Str\"omgren spheres.  
N30B-E is about 7$''$ (1.7 pc) in diameter and shows a bright spot 
at its northern edge, while N30B-W is 6$''$ (1.45 pc) across and has
a rather uniform surface brightness.

\subsection{Densities and Ionization of the \hii\ Regions in N30B}

The $HST$ WFPC2 \ha\ image, with stellar emission excised, can be 
used to estimate the \ha\ luminosities and required ionizing fluxes
of the \hii\ regions in N30B.
We have made two measurements which differ in their assumption as to
whether the halo is a reflection nebula or part of the \hii\ regions.
If the halo around the \hii\ regions is an envelope reflecting the
light from Hen S22, a background component similar to the halo should 
be subtracted.
The resultant integrated \ha\ fluxes for the 7$''$ and 6$''$
Str\"omgren spheres of N30B-E and N30B-W are 
$(5.0\pm0.5)\times10^{-13}$ ergs~cm$^{-2}$~s$^{-1}$ and 
$(3.3\pm0.4)\times10^{-13}$ ergs~cm$^{-2}$~s$^{-1}$, respectively. 
These fluxes have been corrected for the filter transmission and 
\nii\ contamination.
If the halo is ionized, a low background exterior to N30B should be
used and the integrated fluxes of the \hii\ regions including the 
halo would be about 80\% larger; however, as we show in \S3.3 this 
is not the case, so these large fluxes will not be used in further 
discussions.
Adopting the extinction of D106-10 in N30B-W, $E(B-V)$ = 0.14 
\citep{oey96a}, and a distance of 50 kpc, we derive \ha\ luminosities of 
$(2.0\pm0.2)\times10^{35}$ ergs~s$^{-1}$ and 
$(1.4\pm0.2)\times10^{35}$ ergs~s$^{-1}$ and required ionizing fluxes of 
$(1.7\pm0.2)\times10^{47}$ photons~s$^{-1}$ 
and $(1.1\pm0.2)\times10^{47}$ photons~s$^{-1}$ for the Str\"omgren 
spheres of N30B-E and N30B-W, respectively.
Using the average surface brightness and adopting an emitting
path length similar to the diameter of the \hii\ region, we have
further derived the rms densities of N30B-E and N30B-W, both  
$\sim$75 H-atoms cm$^{-3}$.

To examine the stellar content and to estimate the available 
ionizing fluxes in N30B, we have carried out photometry
for stars brighter than $V$ = 20, using the WFPC2 $B$ and 
$V$ images.
Such bright stars are expected to be of early types, so it is 
possible to apply constant zeropoints to transform the WFPC2
$B$ and $V$ to Johnson $B$ and $V$.
The color-magnitude diagram of Johnson $V$ versus $B-V$ is 
presented in Figure 3.
Using the colors and magnitudes and assuming that these
stars are still on the main sequence, we can deredden the 
stars and estimate their spectral types.
This is easily done by projecting the position of a star
in the color-magnitude diagram along the dereddening direction
back to the main sequence.
The photometric measurements and estimated spectral types are 
given in Table 2.  
Most of these stars have also been studied by \citet{oey96b} 
using ground-based observations.
The differences between her results and the new WFPC2 results
are also given in Table 2 for comparison.
Excellent agreements in photometry are found for stars in isolation,
but the ground-based measurements for faint stars in crowded regions 
may have errors as large as $\Delta V \sim$ 1 mag.

In N30B and its vicinity, stars brighter than $V$ = 20 are all 
B stars.
Ionizing fluxes from early-type stars have been given by 
\citet{p73} down to B3 and \citet{SdK97} down to B0.5.
The ionizing fluxes of B0\,V and B0.5\,V stars given by the 
latter are significantly higher than those of the former.
N30B-E contains only one early B star, star \#13 in Table 2; 
the ionizing flux of this B2.5\,V star is $<10^{45}$ photons s$^{-1}$ 
\citep{p73}, two orders of magnitude lower than that required to ionize
N30B-E.
It is possible that the B[e] supergiant Hen S22, of spectral 
type B0--B0.5 \citep{Zetal86}, plays a significant role in the 
photoionization of N30B-E.
N30B-W contains the bright star \#6, which is saturated in the WFPC2 
images, but has been previously classified spectroscopically as 
a B1.5\,V star \citep[= D106-10;][]{oey96a}.
The ionizing flux of a B1.5\,V star given by \citet{p73} 
is 2$\times$10$^{45}$ photons s$^{-1}$, which is also about two 
orders of magnitude lower than the required.
However, extrapolating from the ionizing fluxes of B0\,V and B0.5\,V 
stars given by \citet{SdK97}, we find that the ionizing flux of the 
B1.5\,V star \#6 (D106-10) is sufficient to ionize N30B-W.

\subsection{Relationship between N30B and the Giant Shell DEM\,L\,106}

The relationship between the \hii\ regions in N30B and the 
surrounding giant shell DEM\,L\,106 may be assessed from their
kinematic properties.
An example of our echelle observations is illustrated in Figures 
4a and 4b in two stretches to show the \ha\ and \nii\ $\lambda$6548 
lines from the E-W slit centered near the star \#6 in N30B-W.
The vertical axis is spatial with east on top, and the horizontal axis
is spectral with wavelength increasing to the right.
For convenience, we have converted wavelengths to heliocentric
velocities of the \ha\ line and marked them along the horizontal
axis in Figure 4.
(All radial velocities in this paper are heliocentric.)
The narrowest lines with uniform surface brightness along the slit
are telluric \ha\ and OH lines, located near 0, $-$425, and 
+660 \kms.
The nebular \ha\ and \nii\ $\lambda$6548 lines can be found at 
+300 and $-$380 \kms, respectively.
The \ha\ line in Figures 4a and 4b shows a bright, narrow component 
and a faint, broad component. 
The spatial extent of the broad component coincides with that of
N30B's halo; this component will be discussed in \S3.3.
The narrow \ha\ component arises from ionized gas.
The two \ha\ emission peaks near the center of Figure 4a correspond 
to the \hii\ regions N30B-E and N30B-W, while the fainter \ha\ emission 
extending throughout the entire slit length corresponds to the
giant shell in DEM\,L\,106.
No line splitting is detected in \ha\ or \nii\ for either N30B
or DEM\,L\,106.

To analyze the kinematics of ionized gas quantitatively, we have 
extracted velocity profiles of the \ha\ and \nii\ lines along
each slit, made Gaussian fits, and plotted the results in Figure 5.
The fitted velocities are plotted along the slit positions,
with \ha\ in filled symbols and \nii\ in open symbols;
the fitted FWHM of the \ha\ line is plotted as the ``error bar" 
at each position, except where data points are densely packed.
The observed FWHM of the \ha\ line is the quadratic sum of the 
instrumental FWHM (13.5 \kms), the thermal FWHM (21.4 \kms\ at 
a temperature of 10$^4$ K), and the turbulent FWHM.
Thus the observed \ha\ FWHM of $\sim$26 \kms\ for the \hii\ regions 
in N30B implies a turbulent FWHM of $\sim$6 \kms.
This small internal turbulence is consistent with the expectation for
\hii\ regions ionized by early B-type stars, because they lack strong 
fast stellar winds.
The giant shell in DEM\,L\,106 has a larger observed FWHM in \ha,
35--40 \kms, implying a turbulent FWHM of 24--31 \kms.
The lack of line splitting suggest that this large turbulent FWHM
is dominated by turbulence instead of a systematic shell expansion.

Further kinematic information is provided by the velocity variations 
shown in Figure 5.
The average heliocentric velocity of DEM\,L\,106 is $\sim$ 300 \kms.
Gradual velocity excursions of up to $\pm$10 \kms\ from the average 
velocity are seen.
Along the NS slit centered on Hen S22, a systematic blue-shift of $-$10
\kms\ is seen in the \nii\ $\lambda$6583 line. (The nebular \ha\ line 
is too contaminated by the stellar emission to be measured near the star.)
These blue-shifts provide the only evidence that the internal motion of
DEM\,L\,106 does contain a component of systematic expansion at a level
of $\sim$10 \kms.
The kinematics of DEM\,L\,106 show that this giant shell, 
formed by Hen S22 and the other massive stars in the OB association 
LH38 \citep{LH70}, is by no means a simple expanding shell with a large 
central cavity.
The interior of DEM\,L\,106 may still contain a significant amount of 
gas in the form of dense cloudlets, and the dynamical interactions
between the fast stellar winds and the cloudlets contribute to the
large turbulent FWHM observed in DEM\,L\,106.

The systemic velocities of the \hii\ regions in N30B are similar to
that of DEM\,L\,106, suggesting that N30B is physically associated 
with DEM\,L\,106.
The gas in the vicinity of N30B and especially between its two \hii\ 
regions shows blue-shifts of $-$5 to $-$10 \kms\ from the systemic 
velocity (see Figure 5).
Low-velocity flows can be produced by an evaporation off the 
ionization fronts in \hii\ regions or dynamic interactions 
between stellar winds and \hii\ regions.
In an evaporation flow, the gas projected outside the \hii\ disk 
would have motions perpendicular to the line of sight and show no 
offsets from the systemic velocity.
In a dynamic interaction scenario, if the stellar wind source 
is on the far side of an \hii\ region, the ablation of the \hii\ 
region by the stellar wind would produce a gas flow toward us.
The ubiquitous blue-shifts observed in the vicinity of N30B
indicate that the dynamic interaction between the stellar wind
of Hen S22 and the \hii\ regions in N30B is responsible for 
these low-velocity gas flow.
N30B is probably one of the dense clouds within DEM\,L\,106; 
its visibility is attributed to the early-B stars formed in 
its interior.

\subsection{Spectra and Physical Nature of the Halo of N30B}

We initially speculated that the bow-shock-like halo around 
N30B was produced by interactions between the fast polar wind 
of Hen S22 and the \hii\ regions, and expected to see 
corresponding supersonic kinematic features within the spatial
extent of the halo.
The \ha+\nii\ echellogram of N30B in Figure 4 shows that the 
only kinematic feature that matches the spatial extent of the halo
is a prominent, broad \ha\ emission component.
The continuous distribution of this broad component over 18$''$, 
from the eastern edge of N30B-E's halo to the western edge of 
N30B-W's halo, indicates that the halo envelopes the \hii\ regions
of N30B.
The broad \ha\ emission is centered near 300 \kms\ and detected 
over a velocity range greater than 1000 \kms.
Such high velocities obviously cannot be accelerated by the wind 
of Hen S22; furthermore, such strong shocks would have produced 
more dramatic dynamic effects, such as X-ray emission, which
is not seen \citep{chuetal95}.  

Large velocity widths ($>$ 1000 \kms) have been observed in other
\hii\ regions, for example, the Carina Nebula in our Galaxy and 
the giant \hii\ region NGC\,5471 in M101.
The broad hydrogen Balmer line emission in the Car II region in 
the Carina Nebula has been attributed to the dust-scattered Balmer 
emission of $\eta$ Carinae and is detected at as far as 5 pc from 
$\eta$ Car \citep{LM86}.
The broad \ha\ and \nii\ lines in NGC\,5471 have been suggested
to be powered by supernova explosions and fast stellar winds
\citep{cvc90}.
Clearly, N30B cannot be associated with supernova explosions,
since no X-ray emission has been detected from this region
\citep{chuetal95}.
Furthermore, N30B does not show broad \nii\ lines as are seen in
NGC\,5471, and therefore it should not have the same physical nature 
as the broad-line region in NGC\,5471.
We will thus focus on whether the broad \ha\ line from the halo of N30B 
may be dust-scattered \ha\ emission of the B[e] supergiant Hen S22.

To compare the broad \ha\ emission from the halo of N30B to that 
of Hen S22, we display the echellogram and \ha\ line profiles
of Hen S22 in Figure 4c--d, below the echellogram of N30B.
While they show qualitatively similar profiles with an absorption
component blue-shifted from the centroid of the broad \ha\ emission,
the widths and depths of the absorption are different. 
A similar behavior is observed in the reflection nebula Car II 
near $\eta$ Car.
\citet{LM86} extracted \ha\ line profiles from six positions in 
Car II; these broad profiles (in their Figure 3) all show an 
absorption component on the blue side of the nebular line, but the 
absorption strength decreases radially from $\eta$ Car.
\citet{LM86} suggested that the spatial variations in the
dust-scattered \ha\ line in Car II reflected the temporal
variability of the \ha\ profile of $\eta$ Car \citep{Retal84}.
However, after monitoring the \ha\ profiles of $\eta$ Car and
Car II from 1985 to 1997, \citet{Betal98} show that the
the absorption feature in the reflected \ha\ line is spatially
variable across Car II but temporally invariable over the
12-yr span, and conclude that the reflection nebula is located
within a 46$^\circ$-wide cone of light from $\eta$ Car along the 
axis of its bipolar Homunculus nebula.
We suggest that the broad \ha\ line from N30B's halo is also 
dust-scattered \ha\ emission from a nearby \ha-emitting star,
Hen S22 in this case.
The position-dependence of the broad \ha\ line in N30B is also
caused by different viewing angles, instead of a temporal 
variability of Hen S22, as discussed further in \S3.5.

The reflection nebula around N30B is not unique in the vicinity 
of Hen S22.  Our echelle observations also detected broad \ha\ 
line profiles in the nebulosity at 20$''$ east of N30B, at a 
much fainter level.  
The bow-shock-like morphology of the reflection nebula around 
N30B is also not unique among reflection nebulae, as a similar 
nebula has been observed around the B2\,V star LSS\,3027 in our 
Galaxy \citep{chu83}.
In both cases, the bow-shock-like structure curves with the apex 
toward an OB association, LH\,38 for N30B \citep{oey96a} and 
Stock 16 for LSS\,3027 \citep{chu83}.
The relative nebular and stellar positions are highly suggestive 
that the stellar radiation and winds from the OB association
are responsible for the bow-shock-like morphology of the
reflection nebula.
For N30B, the dynamical interaction is further supported by the 
small blue-shift, $-$5 to $-$10 \kms, of the ionized gas in the
vicinity of the two \hii\ regions.
This flow velocity, $-$5 to $-$10 \kms, is subsonic for 10$^4$ K 
ionized gas but supersonic for cold neutral gas.
Since the reflective envelope of N30B contains dust and 
mostly neutral gas, flow velocities of 5--10 \kms\ would be 
supersonic; thus the morphology of N30B's halo may indeed be 
produced by a bow shock.

\subsection{Geometry of the Reflection}

The nature of N30B's halo as a reflection nebula is supported by
the WFPC2 $B$, $V$, and $I$ images in Figure 2.
These three broad-band images show similar morphologies among one 
another, with almost uniform surface brightness across N30B.
Hardly any enhancement is seen over the Str\"omgren spheres where 
bright \ha\ emission is present, indicating that the contribution 
of nebular line emission from the \hii\ regions is minimal compared
to the scattered stellar continuum and line emission.
Therefore, we may use the observed broad-band surface brightness of 
N30B's reflection nebula and a few simple assumptions to determine 
the likely position of N30B relative to the B[e] star Hen S22.
First, we assume a single value for the spherical albedo of every 
dust grain.  We also assume that the reflection nebula is optically
thick, i.e., every photon from Hen S22 incident upon the reflection 
nebula is either absorbed or scattered.  It has been shown that 
dust scattering has an angular dependence and follows the phase 
function
$
\Phi(\alpha) = \frac{\gamma(1 - g^2)}{4\pi}(1 + g^2 - 2g \cos
\alpha)^{-3/2},
$
where $\alpha$ is the deviation of the photon from the
forward direction, $\gamma$ is the spherical albedo of the dust
grains, and $g$ is a measurement of the asymmetry of the phase 
function as given by the expression, 
$\gamma g=\int\Phi(\alpha)\cos \alpha~d\omega$ \citep{HG41}.
We have adopted dust grain 
properties reported by \citet{LW76} for wavelengths longward of 
3000 \AA, $\gamma = 0.7$ and $g = 0.60$.  
As shown in Figure 6, the scattering angle $\alpha$ is related to 
the projected distance $d$ from Hen S22 to the reflection nebula 
by $\sin \alpha = 5~{\rm pc} / d$.  
Using the phase function and the brightness of Hen S22,
we can calculate the surface brightness of the reflection nebula
expected for a given geometry of the reflection.

The broad-band photometry of Hen S22 from 1957 to 1991 show
stable brightnesses and colors:  $V = 11.80\pm0.05$
and $B-V$ = +0.28 \citep{Zetal86,Zetal96}.
The reddening of Hen S22, $E(B-V)$ = 0.25--0.30, is higher 
than those of the other stars in DEM\,L\,106 because of its 
edge-on disk \citep{Zetal86}.
The broad-band photometry of Hen S22 made by 
\citet{oey96a,oey96b} in 1994 shows a similar $V = 11.879\pm0.004$, 
but a higher $E(B-V)$ = 0.46.
We have adopted an $E(B-V)$ of 0.25 to deredden the Johnson 
$B$ magnitude of Hen S22, and converted it to the WFPC2 $B$ 
($F439W$) magnitude.
We then calculate the expected surface brightness of the 
reflection nebula in the WFPC2 $B$ band for a range of 
scattering angles.
To compare these predictions to observations, a reddening 
appropriate for the reflection nebula needs to be applied.
As the reflection nebula is on the surface of N30B, its 
extinction should be no greater than the minimum extinction
found among stars in N30B, $E(B-V) \sim 0.06$.
We have applied this amount of extinction to the expected
surface brightness of the reflection nebula, and the results
are presented in Table~3.  
The maximum expected surface brightness comes at a scattering 
angle of $\sim$40$^{\circ}$, which corresponds to a distance 
of $\sim$8~pc between Hen S22 and the reflection nebula.  
We have measured an average surface brightness of 
1.6~$\times$~10$^{-14}$~ergs~s$^{-1}$~cm$^{-2}$~arcsec$^{-2}$
for N30B's reflection nebula in the  WFPC $B$ band.
This observed surface brightness is 20\% higher than the
maximum expected surface brightness.
Considering that our assumptions and calculation are crude,
the observed and expected surface brightness of the reflection
nebula are in remarkably good agreement. 
As discussed in \S3.5, the reflection nebula is above the equatorial 
disk and sees a less reddened Hen S22, the expected surface 
brightness of the reflection nebula is thus higher and would be
in even better agreement with the observations.
We conclude that the reflection nebula around N30B is consistent 
with forward scattering by the dust grains, i.e., N30B is on the 
near side of Hen S22.  
The most likely position of N30B is $\sim$8 pc from Hen S22 and 
the scattering angle $\alpha$ is $\sim$40$^\circ$.

\subsection{Periscopic Views of the Polar and Disk Winds of Hen S22}

The broad \ha\ line profiles from the reflection nebula around 
N30B not only change with position (Figure 7) but also differ 
from that of Hen S22 (Figure 4).
Adopting the most likely geometry of reflection suggested in \S3.4,
$d \sim$ 8 pc and $\alpha \sim 40^\circ$, the light travel time from
Hen S22 to us via the reflection nebula is 6 yrs longer than the
direct travel time.
Thus the spectra of the reflection nebula should be compared to the
spectra of Hen S22 taken 6 yrs earlier; unfortunately none are
vailable with sufficient resolution.

Spectra of Hen S22 available in the literature were taken from
the late 1960's to early 1980's.
While these observations yield a consistent systemic velocity
of Hen S22, $293\pm3$ \kms\ \citep{Zetal86}, different
blue-shifts (from the systemic velocity) of the absorption component
in the hydrogen Balmer lines have been reported:
$-$70 \kms\ derived from the H8 to H31 lines observed with a 
reciprocal dispersion of 20 \AA\ mm$^{-1}$ in the years 1968--1975 
\citep{M78}, $-$87 \kms\ from the \ha\ line observed with 
38 \AA\ mm$^{-1}$ in 1979, and $-$40 \kms\ from the H$\beta$,
H$\gamma$, and H$\delta$ lines observed with 20 \AA\ mm$^{-1}$ 
in 1982 \citep{Zetal86}.  
Since these measurements were made for different Balmer lines with
different spectral dispersion at different epochs, it is not clear 
whether the differences in the blue-shifts of the absorption component 
are caused by atmospheric effects (low and high Balmer lines have 
different profiles) or temporal variations in the stellar wind.
Our echelle observations of Hen S22 were made in 2000 January and 
December with a reciprocal dispersion of 3.4 \AA\ mm$^{-1}$.
These two spectra show almost identical \ha\ profiles, with
the absorption component detected at 260 \kms.
Relative to a systemic velocity of 293 \kms, the blue-shift of 
our \ha\ absorption is $-$33 \kms.
To compare our \ha\ profiles to that reported by \citet{Zetal86},
we have convolved our spectra with a Gaussian with $\sigma$ = 32 
\kms, but we find that the overall \ha\ profiles do not match that 
presented in their Figure 5. 
This mismatch might be caused by an incorrect wavelength scale in
their Figure 5, where the displacement of the absorption dip 
from the systemic velocity appears to be $<$50 \kms, significantly 
less than what they reported in their text.

From the above comparisons we find no conclusive or convincing 
evidence to indicate that the \ha\ line profile of Hen S22 varied.
Furthermore, the $V$ magnitude and the $B-V$ color of Hen S22 did 
not change for more than 0.1 mag from 1957 to 1996
\citep{Zetal86,Zetal96,oey96b}.
Therefore, we cannot attribute the spatial variations of the broad 
\ha\ line from the refection nebula to hypothetical temporal 
variations of the \ha\ line of Hen S22.

Figure 7 shows smooth spatial variations of the broad \ha\ profile, 
especially the absorption component, across N30B, similar to
what has been observed in Car II \citep{Betal98}.
The explanation of this spatial variation is best illustrated
by the Homunculus nebula of $\eta$ Car, as described and analyzed
in detail by \citet{smith02}. 
The bipolar lobes of the Homunculus are dusty and scatter the
incident starlight of $\eta$ Car; the reflected spectra at higher
latitudes show \ha\ absorption extending to larger blue-shifts.
As the Homunculus is smaller than 1 ly in size, temporal 
variations in the spectrum of $\eta$ Car are negligible over the 
differential light travel time; therefore, the latitude-dependence 
of the \ha\ absorption in the reflected spectra must indicate an 
anisotropic stellar wind from $\eta$ Car.
Hen S22 has been shown to possess both a slow disk wind and a fast 
polar wind \citep{Zetal86}, consequently its spectral properties 
should also be latitude-dependent.
Thus, the variations in the absorption component in the reflected 
\ha\ emission at N30B must be a result of viewing Hen S22 from 
different latitudes, too.
The reflected spectra on the west side of N30B resemble better
the observed \ha\ line of Hen S22, while those on the east side
show the absorption component extending to larger blue-shifts,
almost $-$100 \kms\ from Hen S22's systemic velocity of 293 \kms.
These variations suggest that the east side of N30B views the
wind of Hen S22 from a higher latitude and the west side of N30B
is close to the disk plane.

The absorption component in the \ha\ line of a B[e] star arises 
from the dense, disk component of the wind close to the star, as
illustrated in the cartoon in Figure 7 of \citet{Zetal85}.
The variations of the absorption component in the reflected 
stellar \ha\ line emission across N30B thus offer an invaluable 
opportunity to study the transition from the disk component to 
the polar component of the stellar wind of a B[e] star.
Future high-S/N, high-resolution spectroscopic observations 
and stellar atmospheric modeling are needed for a better
understanding of the stellar wind of the hybrid-spectrum
B[e] supergiant Hen S22.

\section{Summary}

$HST$ WFPC2 \ha\ images reveal a bow-shock-like halo around 
the compact \hii\ regions in N30B on the side that faces the B[e] 
supergiant Hen S22.  Long-slit, high-resolution spectroscopic
observations of the \ha\ line show both narrow and broad 
emission components in N30B.  The narrow component originates
from the ionized gas in the \hii\ regions; its velocity 
variations reveal a gas flow at $-$5 to $-$10 \kms\ relative 
to the \hii\ regions, indicating dynamic interactions with 
the stellar wind from Hen S22.

The broad \ha\ emission component has the same spatial extent as 
the halo of N30B, but its width ($>$1000 \kms) is too high to 
have been produced dynamically without detectable X-ray emission.
The spectral shape of the broad \ha\ emission is similar to that
of Hen S22, both having a blue-shifted absorption component.
We suggest that the halo of N30B is a reflection nebula
scattering the stellar emission of Hen S22.
$HST$ WFPC2 broad-band images show morphologies consistent with 
this interpretation.
The bow-shock-like morphology is probably indeed formed as
a bow shock, as the downstream flow velocity of $-$5 to $-$10
\kms\ is supersonic to the neutral gas in the halo of N30B.

Assuming typical properties of interstellar dust, we have 
modeled the surface brightness of the reflection nebula of
N30B for a range of scattering angles.
We find that the observed surface brightness of the reflection
nebula in the WFPC2 $B$ band is within 20\% from that expected 
from dust-scattering at a distance of $\sim$8 pc from Hen S22 
for a scattering angle of $\sim$40$^\circ$.

The absorption component in the broad \ha\ emission profile 
varies across the reflection nebula of N30B as a result of 
different viewing angles.
The west side of N30B lies closer to the equatorial
disk plane of Hen S22, while the east side of N30B is at 
higher latitudes.
We suggest that the anisotropic wind of the B[e] supergiant 
Hen S22 can be studied by modeling the broad \ha\ line profile 
and the blue-shifted absorption component, taking into account 
the 3-dimensional geometry.
This unique opportunity could reveal important insights on the
physical conditions and evolution of these enigmatic massive stars.

\acknowledgments 
We thank the anonymous referee for pointing out the existence
of monitoring observations of $\eta$ Car and its reflection 
nebula Car II.  We also thank Nathan Smith and Nolan Walborn 
for useful discussions, Jayanne English for suggesting 
DEM\,L\,106 to the Hubble Heritage Project, and Monica Shaw for 
assistance in initial data analyses.  This research was supported 
by the {\it Hubble Space Telescope} grant STI6698.01-95A.

\clearpage

\begin{table}[h]
\caption[junk]{Journal of Echelle Observations}
\vskip 10pt
\renewcommand{\footnoterule}{}
\begin{minipage}{5.75in}
\begin{tabular}{lccccc}
\hline \hline
 Number&  Slit             & Slit    &   Exposure  &   Date of \\
       &  Center           & Orient. &   (s)     &  Observation  \\
\hline
   1   &  Hen~S22          &  NS     &    600      &   2000 Jan 22 \\
   2   &  Hen~S22          &  NS     &    300      &   2000 Jan 22 \\
   3   & 19$''$S of Hen~S22&  NS     &    300      &   2000 Jan 22 \\
   4   &  Hen~S22          &  EW     &    300      &   2000 Jan 22 \\
   5   & 19$''$S of Hen~S22&  EW     &    300      &   2000 Jan 22 \\
   6   & 23$''$S of Hen~S22&  EW     &    300      &   2000 Jan 22 \\
   7   & 27$''$S of Hen~S22&  EW     &    300      &   2000 Jan 22 \\
   8   & Hen~S22           &  EW     &    ~30      &   2000 Dec ~7 \\
   9   & near D106-10      &  EW     &    600      &   2000 Dec ~7 \\
\hline
\end{tabular}
\end{minipage}
\end{table}

\clearpage

\vskip 10pt
\begin{table}[h]
\begin{center}
\caption{Photometric Measurements of Stars in the Vicinity of N30B}
\begin{tabular}{rccccccl}
\tableline\tableline
Star & R.A.~(J2000)	& Dec.~(J2000)	   & $B$      & $V$     &
$B-V$ & $V-V_{\rm Oey}$\,\tablenotemark{a} & Spectral \\
No.  & (h m s)          & ($^\circ~'~''$)  & (mag)   & (mag)  &
(mag) &  (mag)          & Type\,\tablenotemark{b} \\
\tableline

 1 & 5 13 49.45 &$-67$ 27 19.0 & 18.73$\pm$0.01& 18.76$\pm$0.01& $-0.03\pm$0.02~~~& $-0.01~~$& B9    \\
 2 & 5 13 50.06 &$-67$ 27 29.4 & 18.29$\pm$0.01& 18.26$\pm$0.01&  0.03$\pm$0.01& $-0.08~~$& B6.5  \\
 3 & 5 13 50.45 &$-67$ 27 22.4 & 16.11$\pm$0.00& 16.26$\pm$0.00& $-0.15\pm$0.01~~~& 0.05& B2    \\
 4 & 5 13 50.52 &$-67$ 27 18.1 & 17.64$\pm$0.01& 17.69$\pm$0.01& $-0.05\pm$0.01~~~& 0.65& B5    \\
 5 & 5 13 50.68 &$-67$ 27 18.5 & 17.48$\pm$0.01& 17.59$\pm$0.01& $-0.11\pm$0.01~~& ...& B5.5  \\
 6 & 5 13 50.76 &$-67$ 27 18.2 & saturated   & saturated     & ... &
   ... & B1.5\tablenotemark{c} \\
 7 & 5 13 50.82 &$-67$ 27 20.0 & 18.40$\pm$0.01& 18.45$\pm$0.01& $-0.05\pm$0.02~~~& 0.81& B8    \\
 8 & 5 13 51.11 &$-67$ 27 20.4 & 18.31$\pm$0.01& 18.30$\pm$0.01&  0.01$\pm$0.02& 0.63& B7    \\
 9 & 5 13 51.34 &$-67$ 27 22.0 & 17.80$\pm$0.01& 17.70$\pm$0.01&  0.10$\pm$0.01&
    0.62& 2$\times$B7\tablenotemark{d}\\
10 & 5 13 51.49 &$-67$ 27 20.1 & 18.24$\pm$0.01& 18.19$\pm$0.01&  0.05$\pm$0.01& 0.70& B6    \\
11 & 5 13 51.55 &$-67$ 27 17.4 & 18.91$\pm$0.02& 18.83$\pm$0.01&  0.08$\pm$0.02& 1.02& B8    \\
12 & 5 13 51.59 &$-67$ 27 21.4 & 19.39$\pm$0.03& 19.14$\pm$0.02&  0.26$\pm$0.03& 1.26& B7    \\
13 & 5 13 51.90 &$-67$ 27 18.7 & 17.85$\pm$0.01& 17.72$\pm$0.01&  0.13$\pm$0.01& 0.42& B2.5  \\
14 & 5 13 52.04 &$-67$ 27 04.7 & 17.61$\pm$0.01& 17.71$\pm$0.01& $-0.10\pm$0.01~~~& 0.04& B6    \\
15 & 5 13 52.67 &$-67$ 27 32.2 & 17.79$\pm$0.01& 17.82$\pm$0.01& $-0.03\pm$0.01~~~& 0.04& B5.5  \\
16 & 5 13 53.22 &$-67$ 27 11.2 & 16.79$\pm$0.01& 16.88$\pm$0.01& $-0.09\pm$0.01~~~& 0.03& B2.5  \\

\tableline
\end{tabular}
\tablenotetext{a}{$V_{\rm Oey}$ is from \citet{oey96b}.}
\tablenotetext{b}{The spectral types are estimated from the color and 
  magnitude of the stars assuming that they are on the main sequence.}
\tablenotetext{c}{This star was cataloged as D106-10 and spectroscopically
  classified as B1.5\,V by \citet{oey96a}.}
\tablenotetext{d}{The photometric measurements include two stars
  of similar brightness.  The spectral type is estimated assuming
  two identical stars.}
\end{center}
\end{table}

\clearpage

\vskip 10pt
\begin{table}[h]
\caption{Expected Surface Brightness of the Reflection
Nebula around N30B \label{tblSB}}

\begin{tabular}{ccc}
\hline \hline
Scattering &          & Surface \\
Angle      & Distance & Brightness \\
(deg.)     & (pc)     & (ergs s$^{-1}$ cm$^{-2}$ arcsec$^{-2}$) \\
\hline
120 & 5.8  & 2.4 $\times$ 10$^{-15}$ \\
110 & 5.3  & 3.4 $\times$ 10$^{-15}$ \\
100 & 5.1  & 4.4 $\times$ 10$^{-15}$ \\
90  & 5.0  & 5.6 $\times$ 10$^{-15}$ \\
80  & 5.1  & 7.0 $\times$ 10$^{-15}$ \\
70  & 5.3  & 8.5 $\times$ 10$^{-15}$ \\
60  & 5.8  & 1.0 $\times$ 10$^{-15}$ \\
50  & 6.5  & 1.2 $\times$ 10$^{-15}$ \\
40  & 7.8  & 1.3 $\times$ 10$^{-14}$ \\
30  & 10.0 & 1.2 $\times$ 10$^{-14}$ \\
20  & 14.6 & 9.3 $\times$ 10$^{-15}$ \\
\hline
\end{tabular}
\end{table}


\clearpage
\begin{figure}
\begin{center} {\large \bf Figure Captions} \end{center}

\caption{(a) CTIO Curtis Schmidt \ha\ image of DEM\,L\,106.
 (b) $HST$ WFPC2 \ha\ image of Hen S22 and N30B.  The echelle slit
  positions centered at 19$''$S, 23$''$S, and 27$''$S of Hen S22 
  are marked by brackets.}
\end{figure}

\begin{figure}
\caption{$HST$ WFPC2 images of N30B in (a) \ha, (b) [\ion{O}{3}],
(c) [\ion{S}{2}], (d) $B$, (e) $V$, and (f) $I$.  Stars marked in
 (e) are brighter than  $V$ = 20 mag and have photometric measurements 
 given in Table 2.}
\label{fig:hst}
\end{figure}

\begin{figure}
\caption{The color-magnitude diagram of $V$ versus $B-V$ for stars
 in the vicinity of N30B.  The 
 identification of the stars are marked in Figure 2e and the photometric
 measurements are listed in Table 2.  The location of the main sequence 
 is plotted for a distance of 50 kpc, or a distance modulus of 18.5 mag.
 A reddening vector of $A_V$ = 1 mag is also plotted.}
\end{figure}

\begin{figure}
\caption{(a) and (b) Echellogram of the \ha\ and \nii\ $\lambda$6547
 lines for the E-W slit position centered near the star D106-10
  (star \#6 in Figure 2e) presented in two stretches to show the 
  bright and faint features.  
  (c) Echellogram of the \ha\ line of  Hen S22 observed
  in 2000 December.  (d) \ha\ line profiles of Hen S22 observed in 2000
  January and December.  
  The horizontal axis is heliocentric velocity of the \ha\ line.
  See text for more explanations.}
\end{figure}

\begin{figure}
\caption{Position-velocity plots along four echelle slit positions.
  The radial velocities are heliocentric. 
  The zero point and direction of the position axis are given on
  the top margin of each plot.
  The filled symbols are \ha\ measurements and open symbols
  [\ion{N}{2}] measurements.  The FWHMs of the
  \ha\ line are plotted as error bars.  The E-W slit position
  centered at 19$''$ south of Hen S22 is almost coincident with
  that centered near D106-10, so their measurements are plotted
  in the same panel with the former plotted in squares and the
  latter in triangles.}
\end{figure}

\begin{figure}
\caption{Geometry of the reflection.}
\end{figure}

\begin{figure}
\caption{\ha\ echellogram and line profiles of N30B along an E-W slit
  centered near the star D106-10 in N30B-W.   The \ha\ line profiles
  are extracted from apertures p1, p2, p3, and p4 marked in
  the top panel.  The horizontal axis is in heliocentric 
  velocity of the \ha\ line.  The narrow lines at $V_{\rm Hel}$ = 
  $-$3 and +660 \kms\ are telluric \ha\ and OH lines, respectively.}
\end{figure}

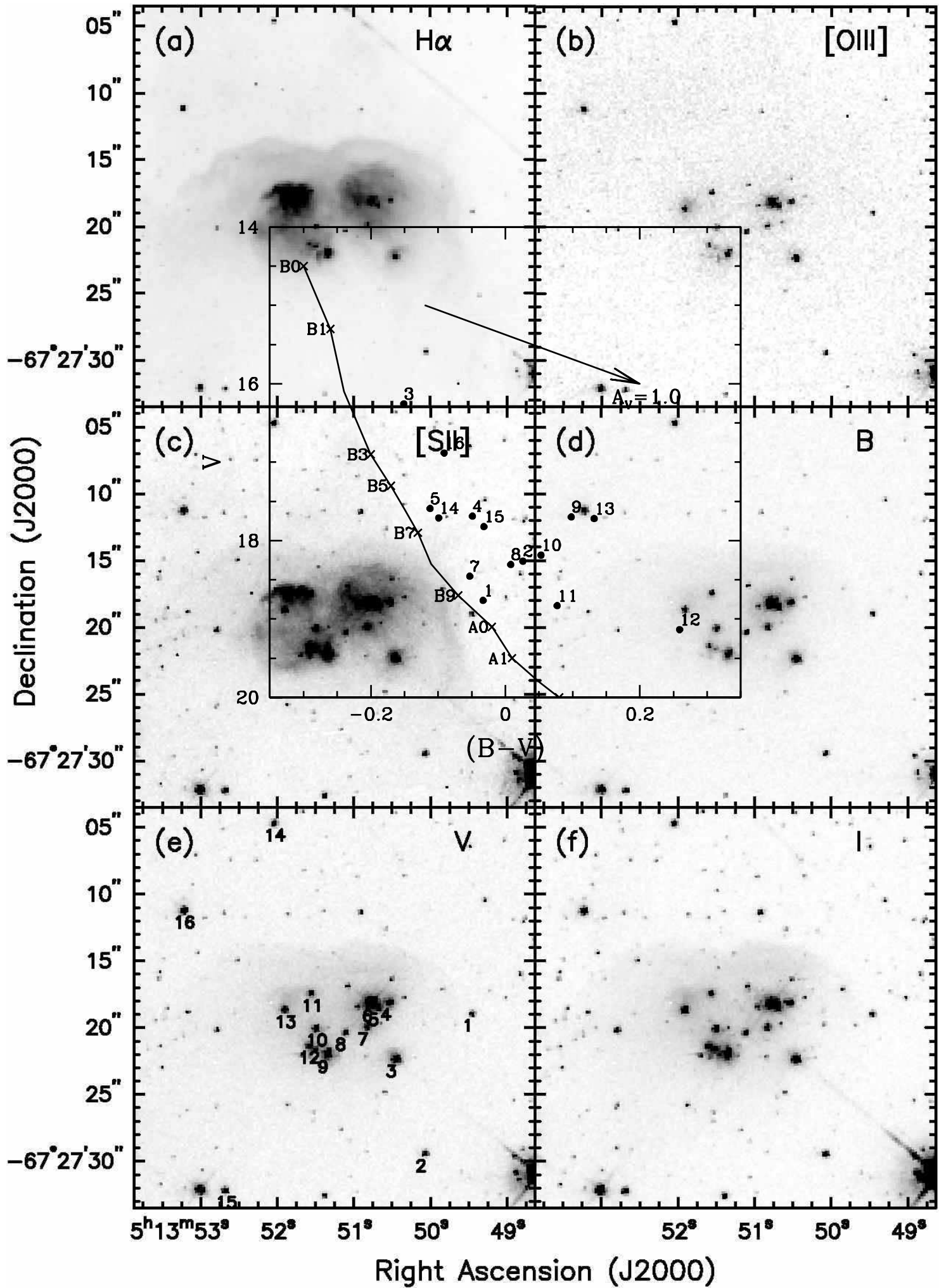
\end{document}